# Transient optical gratings for pulsed ionizing radiation studies

W.K. Fullagar[a,b,*], D.M. Paganin[a], C.J. Hall[c]

[a] School of Physics, Monash University, VIC 3800, Australia
[b] Australian Research Council, Centre of Excellence for Coherent X-Ray Science, Australia
[c] Australian Synchrotron, VIC 3800, Australia

*Abstract*

Prior to the invention of holography or lasers, Bragg's X-ray microscope [1, 2] opened the door to optical computation in short-wavelength studies using spatially coherent visible light, including phase retrieval methods. This optical approach lost ground to semiconductor detection and digital computing in the 1960s. Since then, visible optics including spatial light modulators (SLMs), array detectors and femtosecond lasers have become widely available, routinely allowing versatile and computer-interfaced imposition of optical phase, molecular coherent control, and detection. Today, high brilliance X-ray sources begin to offer opportunities for atomic resolution and ultrafast pump-probe studies. Correspondingly, this work considers an overlooked aspect of Bragg's X-ray microscope - the incoherent ionizing radiation to coherent visible (IICV) conversion that is a necessary prerequisite for coherent optical computations. Technologies are suggested that can accomplish this conversion. Approaches to holographic data storage of short wavelength scattered radiation data, and phase retrieval that incorporates optical procedures, are motivated.

PACS codes : 06.60.Jn, 29.40.-n, 78.47.jj, 82.50.Kx

## I. OPPORTUNITIES

X-rays are a form of light whose wavelengths match molecular interatomic spacings. This underpins not only the routine use of X-rays for molecular and small object structure determinations, but also the early realisation that propagation of suitably phased spatially coherent visible light fields can "undo" the X-ray scattering process to obtain greatly magnified optical reconstructions of the invisibly small objects that caused the X-ray scatter. This is the basis of Bragg's X-ray microscope [2], which is a viable approach to the observation of molecular structures

* corresponding author : w_fullagar@hotmail.com

[3-11] using X-rays and other short wavelength radiations [12]. The importance of introducing phase information in the optical reconstruction procedure was recognised at the outset by Bragg. In the case of scattering of optical wavelengths, Gabor's invention of holography [13] was a fundamental step in the retention of phase information. With ionising radiations, the situation is more complicated. At high diffraction angles (momentum transfer) corresponding to high spatial frequencies within the sample and the consequent ability to resolve them, X-ray optics become restrictive. High numerical apertures are not possible, while available devices increasingly entail narrow design parameters, flux losses, and cost, at the same time that disposability of the optic can become necessary if extreme fluxes result in rapid damage. Providing a holographic reference beam for high angle sample scatter might be achieved in other ways, but there remains a further critical consideration, as follows. As the diffraction angle increases, the spatial period of interferometric phase oscillations in a far-field detection plane decreases, eventually becoming comparable to the radiation wavelength. In this high-resolution context, the preservation of phase in conventional holographic recordings requires that radiation event sizes in the recording medium are small compared to the causative photon wavelength. However, this situation does not obtain at electromagnetic wavelengths corresponding to interatomic distances, since as the photon energy increases, the radius of thermalized events also increases. Indeed, the relationship between these parameters [14] indicates that event diameters match the corresponding electromagnetic wavelength at energies ~70 eV (~18 nm) in typical condensed matter, restricting conventional holographic approaches already at this low energy. Despite this smearing of phase information when the spacing of interference fringes becomes comparable to event diameters, the interrogation of radiation-induced events using longer wavelengths (in particular visible light, with wavelength typically ~500 nm) is not precluded. Even at "hard" X-ray energies of ~50 keV (leading to thermalized event diameters ~5 μm in condensed matter [14]), it is quickly appreciated that optical scatter from entities with this large event size will involve scattering angles that are easily experimentally accessible [15]. However, having lost the short wavelength phase information, it is necessary to both determine and re-introduce it in any subsequent reconstruction. Early methods were devised for doing this optically [8, 9], which nowadays would be accomplished through the computer-interfaced combination of lasers, SLMs [16] and optical image transducer technologies [17-19]. Meanwhile, digital X-ray detection and computation technologies have matured to the point of near-universal use, so that Bragg's X-ray microscope lapsed into historical obscurity.

We now face the prospects of brilliant X-ray free electron laser (XFEL) sources [20], lab-based ultrafast laser driven short wavelength radiation sources [21-24] and their extensions [25], as well as increasingly high coherence electron beams. With lasers, computer-interfaced optical imaging capabilities and SLMs already in widespread use, it is timely to re-examine the case for Bragg's X-ray microscopy and related studies, such as X-ray photon correlation spectroscopy (XPCS) [26-28]. Techniques for IICV conversion are a missing link that impedes their contemporary redevelopment. We offer the following motivations for overcoming this impediment. Depending on the chosen approach, the dynamic range of X-ray detection in short-pulse experiments could potentially be dramatically improved, circumventing pileup issues. One can avoid issues related to detector pixelation. Potential exists for detector readout during brief intervals between ultrashort pulses. Effective and affordable lenses, beamsplitters and coherent optical gain media are highly developed for optical wavelengths, with corresponding devices at X-ray wavelengths at a lesser stage of both development and flexibility. Optical laser diffraction-based computation is the natural cousin of monochromatic X-ray diffraction, with the former now enjoying a mature yet extremely widespread and active state of development [29-31]. Optical approaches offer clear paths to new developments, such as optical artificial intelligence possibilities for X-ray phase retrieval [32-34], and 3-dimensional holography [35, 36] to retrieve phase information through transport of intensity concepts [37]. The maturity of ultrafast coherent control at the molecular level [38-40] points to *in-situ* short wavelength molecular structure observations on the same molecular dynamics timescales [41, 42], while ultrafast X ray – optical pump-probe studies would be an integral aspect of sample, detection, and data reduction technologies. The monochromatic approach implicit in this work, though restrictive in its potential for ultrafast work [42, 43], offers interesting extensions to earlier successful optical pump, X-ray probe developments for observing transient molecular structures [44-46]. In short, addressing ionizing radiation to visible, incoherent to coherent conversion, particularly as transient gratings in modern ultrafast pump-probe contexts, opens a new door to the science of short wavelength detection, analysis, data reduction, and through optical holography, data storage and retrieval [47]. Section III additionally suggests a novel and substantially optical approach to X-ray phase retrieval, which may be compared to the substantially digital approaches in common current use.

## II. APPROACHES

Figure 1 sketches one arrangement that performs the essential IICV SLM conversion. Variations are easily realised and may offer advantages [43]. Note the critical role of spatial coherence in the optical field. The use of light emitting phosphors or incoherent observations of metastable color centers through various relaxation channels is thus essentially unrelated, though such studies can be both inspirational and relevant [48]. A wide variety of future developments might be anticipated by the optical detection we prescribe. To accomplish transient grating IICV, several approaches are possible, with the route chosen being dependent on demands of particular applications. Since there is no succinct way to describe every potential application and its pertinent details, the intention here is simply to provide an overview. We particularly indicate the following technologies (see also [43]):

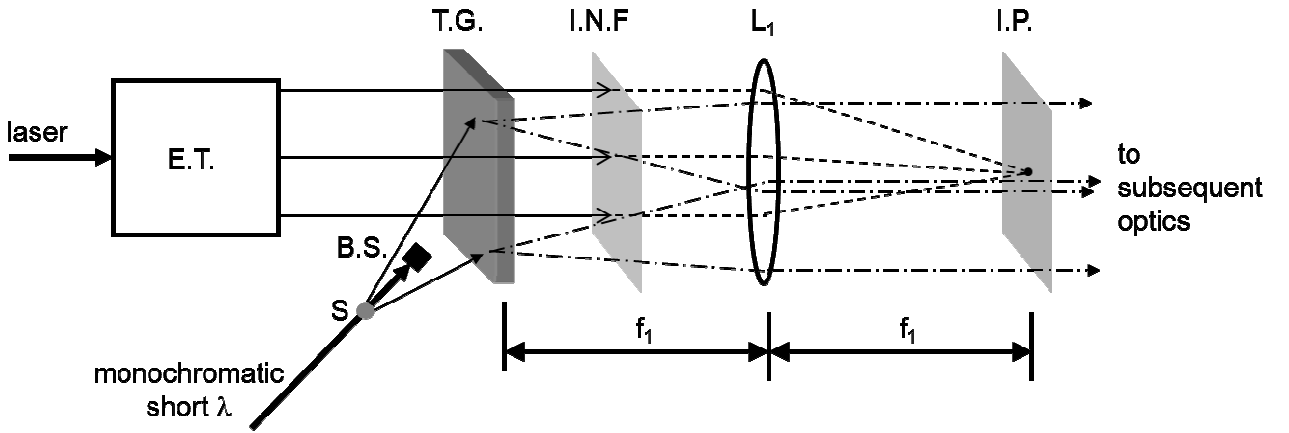

FIG. 1. Pulsed monochromatic short wavelength radiation scatters from a sample S onto a transient grating window TG (BS = beamstop). Within the timescale of the transient grating, radiation-induced scattering centers in TG scatter an expanded and collimated pulsed laser beam (ET = expansion telescope). Unscattered laser light is prevented from further propagation by the (angularly selective) interference notch filter (INF) and/or spatial filtering at the image plane IP. Coherent optical processing can then be performed in subsequent optical stages.

*a) Color centers (including photography)*

Ionizing radiation events can cause optically absorbing features in transparent media (or transparency in opaque media). Dye-based IICV SLMs have been developed in all-optical contexts [49] but apparently not for X-ray or other ionizing radiation, despite widespread use of chemical dosimetry. Optical efficiency of absorption-based diffraction gratings is lower than for phase gratings [50, 51]. On the other hand, transient gratings involving F-centers and solvated electrons in condensed matter have been studied for well over a century, including large volumes of picosecond and femtosecond work in recent decades. They can offer extremely wide limits of temporal latency and exceptional dynamic range (being limited by recombination dynamics and the material's capacity to accommodate free electrons) [52-55]. Optical detectability of single X-ray events is questionable, but with the consolation that optical reconstructions are impossible based on small numbers of scattered ionizing quanta (a large ensemble of events is essential, implying no need for their individual detectability). Chemical imaging procedures incorporating gain (typically photography) include radiation-induced polymerisation [56, 57]. Gain improves sensitivity, but at the expense of available dynamic range and spatial resolution. It also implies a drop in chemical free energy, indicating that starting materials are in a metastable state and that changes are irreversible. The ability to rapidly regenerate the metastable state is then required if used as a transient optical grating. Regeneration could be by optical, electrical, pressure/acoustic, thermal or other means.

*b) Liquid crystal X-ray light valves*

In conjunction with optical polarisers, and for medical X-ray imaging, these devices have been under development for several years [58]. To our knowledge, they have not been developed for IICV purposes, though the technology appears suitable. Charges induced by ionizing radiation in a semiconductor cause local alterations of birefringence in an adjacent thin film of liquid crystal, leading to modulations of optical phase in a probing light field. Foreseeable complications are that the dynamic range for the optical phase modulation is limited by the possible extent of molecular configuration change in the liquid crystal. In addition the kinetic rates of the molecular reorientation processes are slow (typically 0.01 - 0.1 s) when compared to repetition rates in many pump-probe experiments (e.g. kilohertz).

*c) Ferroelectrics*

In ferroelectrics optical birefringence is modulated by local electrical fields, including free charges generated by radiation-induced ionization in the bulk of the material. The effect on a spatially coherent light field is similar to the liquid crystal devices, though here unconstrained by macromolecular reorientation. Thus, through the depth of a ferroelectric crystal, local optical phase shifts can occur to the extent that phase wrapping occurs. The kinetics of the atomic repositioning associated with birefringence are also very different, being dependent on the proximity to the Curie temperature [50] and ultrafast material chemical dynamics [59-61]. An important work in the present IICV context is [62]; we also draw attention to references [63-65]. Acoustic translation and spatiotemporal manipulation of excitonic states in related devices [66] raises the possibility of a level of control comparable to that employed in the readout mechanism of conventional charge coupled devices (CCDs).

*d) Cloud and bubble chambers*

Here, optical detectability of single events is associated with gain in a supersaturated vapor or supercritical liquid, with nucleation points caused by ionizing radiation. Liquid-gas phase boundaries scatter light in both cases. Many theoretical and experimental examinations of the rate and growth mechanisms of the droplets/bubbles have been made [67-69], including in devices driven by pressure waves and ultrasound [70]. The latter devices could be temporally phase locked to pulsed ionizing radiation sources (e.g. FELs) and pulsed interrogation lasers [71-73] to permit control of the optical scatter of growing droplets/bubbles by modulating the amplitude and temporal phase of the acoustic field. Bubble chambers offer short stopping lengths for ionizing radiations, as motivated in Figure 1. On the other hand, relatively long stopping distances in the gaseous medium of cloud chambers opens the possibility of recovering short wavelength phase information as the radiation field propagates away from a scattering sample, through transport of intensity concepts [37]. Experimentally this would involve holographic examination of the cloud density. A potential complication is that, depending on their energy, ejected photoelectrons can have considerable ranges in gases, lowering measurement resolution. Indeed photoelectron trails are mainly responsible for the cloud's visibility, as seen in the beautiful and pioneering X-ray pump, optical probe work of Wilson [74].

### III. FUTURE SCOPE FOR OPTICAL APPROACHES

The sketches of Figures 2 and 3 suggest two possibilities opened by this detection approach, indicating the potential for holographic data storage and adaptive optics approaches to phase retrieval, respectively. (In practice the phase retrieval step may be better done on a recorded hologram, rather than directly on the transient grating itself, since the hologram's purpose is to be a recording of otherwise transient data.)

In Figure 2 the use of a volume hologram is suggested in which one rotation axis of the hologram is coupled to a rotation of the sample. A perpendicular axis is coupled to some other parameter of interest; here we suggest a sample pump-probe delay interval. The reconstructed hologram would then show information pertaining to sample rotation about the first viewing axis, and the sample's temporal evolution about the second axis. Various optical holographic media could prove useful for the purpose (for example [47, 50, 75, 76]) noting limitations on the quantity and fidelity of data that can be stored in this way (for example [34]). In recording, the stepping of the holographic volume's rotation might usefully be synchronised to the pulsed ionizing radiation. It is also evident that spatiotemporal overlap of the sample and reference laser light must occur within the holographic recording volume.

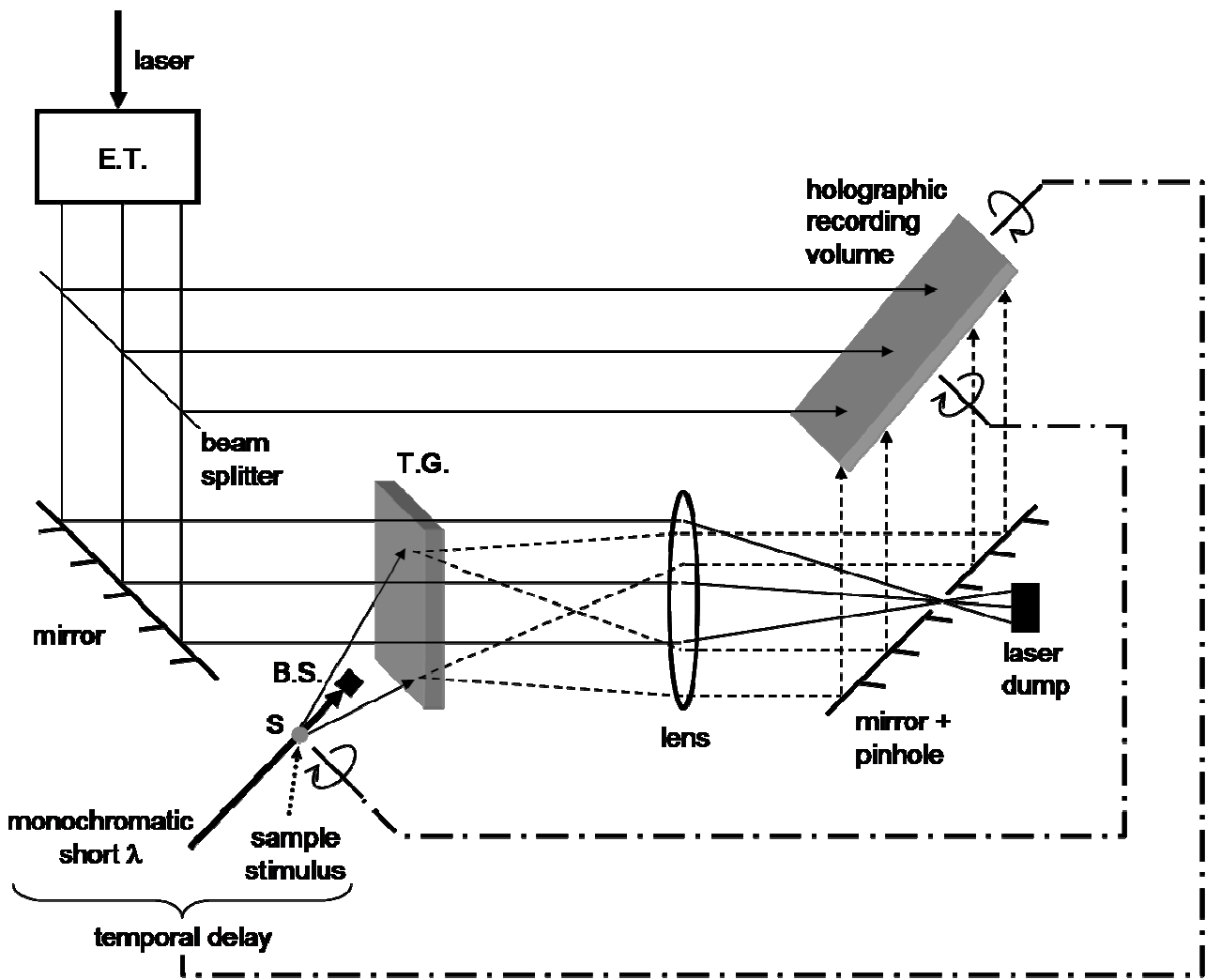

FIG. 2. An extension of Figure 1 indicating the use of a volume hologram for X-ray data storage (see text).

Figure 3 suggests how adaptive optics might be used to assist recovery of short wavelength phase information (arrangements for optical pattern recognition use a comparable configuration [77]). The illustrated approach may be attractive, given that relevant hardware with integrated computer interfacing can be found, for example, in LCD data projectors [78], and that increasingly, open source adaptive optics software is freely available [79]. Here, a folded geometry of the transient grating apparatus is sketched, motivated elsewhere [43], in which the transient grating might consist of a wafer of electro-optic material coated on the ionizing radiation side by a radiation transmissive, optically reflective layer. Lens 1 and lens 2 are shown in a 4f geometry, with the transient grating on one side, the SLM on the other, and the mirror with pinhole in the center. The transient grating and pinhole in mirror on either side of lens 1 are used exactly as in the detection system in [43], though here with the light propagating in the reverse direction. Thus, transmission of laser light through the pinhole will be maximised when applying an optical amplitude and phase field that is the Fourier transform of the short wavelength scatter to the mirror in the plane of the pinhole, in circumstances where multiple scattering is unlikely [43]. The function of the algorithm is to determine and generate this field [80, 81]. Lens 2 is placed between the mirror and the SLM, so that the pattern presented on the SLM is in fact the Fourier transform of the field required in the plane of the mirror. The phase pattern presented on the SLM is thus the double Fourier transform of the short wavelength scattering object; in other words, it is the phase pattern of the short wavelength scattering object itself, as reconstructed by the computer algorithm. This outcome could be stored holographically if desired. The laser wavelength and distances employed will determine the relative spatial scaling of the optical *vs* ionizing radiation phase patterns.

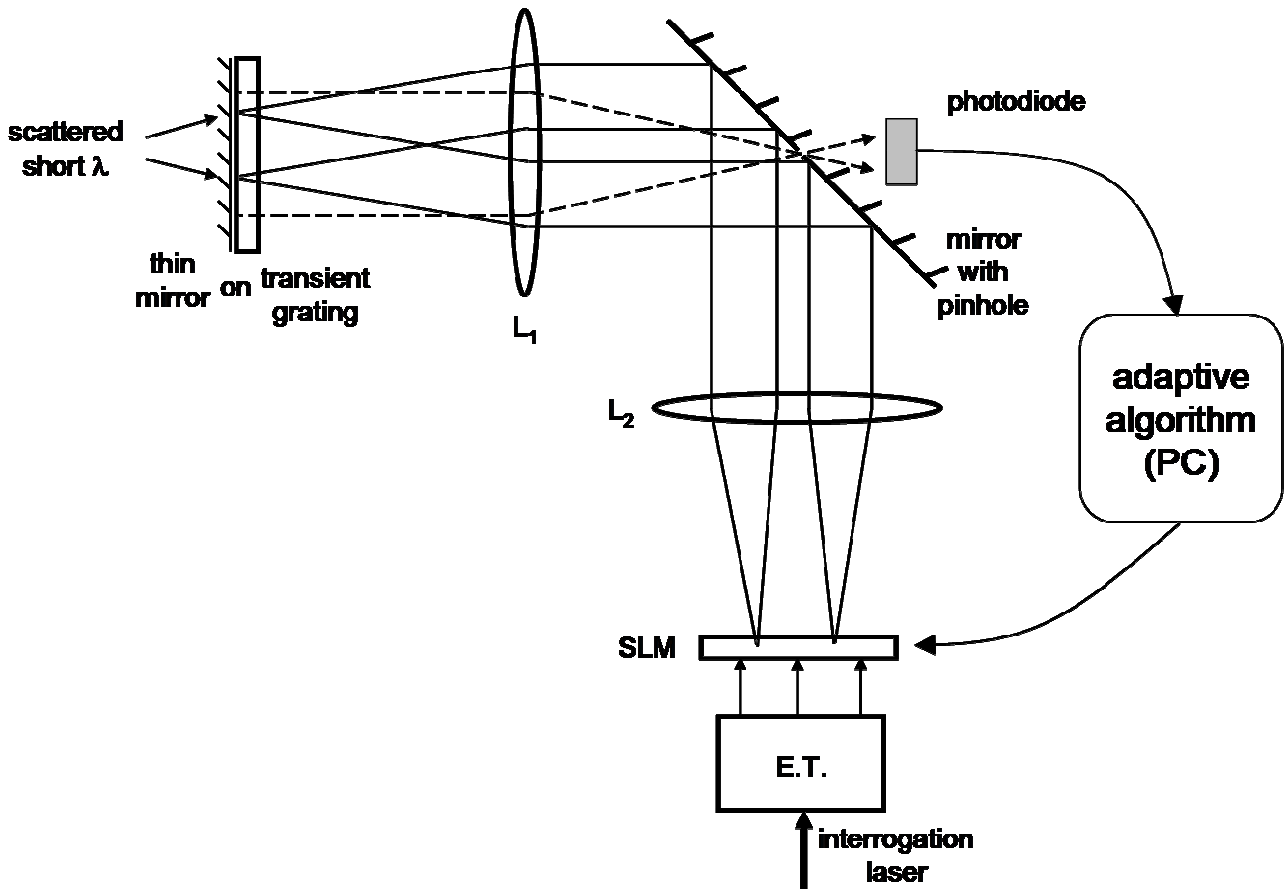

FIG. 3. A folded geometry transient grating, motivating the use of algorithms that adapt the optical wavefield in order to recover phase information in the scattered short λ wavefield (see text), in a manner analogous to guide-star methods used by the astronomical adaptive optics community.

## IV. MOLECULAR STRUCTURE-DYNAMICS

We put aside the phase reconstruction problem for a moment, while noting that the use of monochromatic ionizing radiation is implicit in the schemes above. The observation of molecular and sample structure factors requires their presentation within the narrow Ewald reflection sphere (which, for the particular case of crystal diffraction, corresponds to meeting the Bragg reflection condition). For FEL radiation sources, the product of source size, beam divergence, spectral width and pulse duration is deliberately minimised, to maximise the brilliance and associated spatial and temporal coherence. The consequently dense and narrow phase space occupation corresponds to extremely thin Ewald spheres in reciprocal space for diffraction. Thus only a two dimensional slice of the sample's four-dimensional structure factor gives observable diffraction in any one radiation pulse (samples possess three spatial dimensions and evolve temporally in pump-probe studies). Mapping out this four dimensional situation is then a major undertaking, the more so if the sample is destroyed on a shot-to-shot basis (by either pump or probe radiations). In monochromatic synchrotron-based crystallographic contexts this difficulty has been successfully overcome [45, 46] by making suitable compromises of sample and experimental constraints, including temporal resolution [44]. Similar problems now face FEL pump-probe molecular studies.

One solution is to use a source that is less brilliant by virtue of its broader spectral bandwidth, allowing a continuum of Ewald spheres to simultaneously probe the structure factors of the sample in each radiation shot. This is the philosophy behind successful ultrafast broadband Laue pump-probe crystallography [82, 83], as well as neutron Laue crystallography and many neutron time-of-flight instruments (where approaches are motivated to make best use of available flux) [84]. Beam divergence can be exploited similarly. More generally, there are fundamental reasons to reduce the brilliance of the radiation source (without reducing the basic photon flux), to allow pulsed X-ray observation of structural dynamics in samples. In ultrafast pump-probe contexts, and in conjunction with new detector developments [85], recognition of this situation has for several years formed the rationale for powerful new laboratory-based ultrafast X-ray techniques using laser-generated X-rays [41, 42].

In terms of individual bond breakages in chemistry, time scales of $10^{-14} - 10^{-13}$ s are relevant, corresponding to typical molecular oscillations at infrared frequencies. For snapshots of chemical structure, the use of femtosecond pulses of pump and probe radiation is therefore appropriate and necessary. However, within large, somewhat flexible structures (e.g. proteins) the exact relative positioning of atoms from one molecule to the next (and indeed from one moment to the next) can make it necessary to consider the atoms as being blurred together - only the gross superstructure can be meaningfully deduced from measurements that require examination of multiple macromolecules. Without this localisation property, the chemical crystallographer's usual phase retrieval constraint (that matter is constructed of individual atoms, each with well known scattering properties) cannot be easily applied. Further, when aiming for resolutions of (say) a few nanometers in individual biological molecules, it is typically the case that the secondary and higher levels of structure (e.g. α-helices and β-sheets in proteins) are substantially due to hydrogen bonds, which may be more or less labile [86]. The velocity of phonon propagation via these bonds determines the time scale on which macromolecular structure evolution occurs, and consequently the time scale of short wavelength structural probing needed to observe structural dynamics. The speed of sound in water (~1500 ms$^{-1}$) can serve as a guide to phonon velocities in hydrogen bonded biological materials. For example then, if a resolution of (say) 3 nm is sought in a protein molecule (or seems plausible from an instrumentation perspective), functional dynamics of the molecule would not ordinarily motivate pulses briefer than ~2 ps. In other words, functionally relevant structure changes on a 3 nm length scale would be at best only barely discernible on a 2 ps time scale, excepting perhaps where strongly coupled coherent electronic and nuclear motions are involved [87-90]. Such pulses would in fact be considered rather long in modern ultrafast optics, also XFEL science [20]. At the same time, a biomolecule isolated in vacuum can rarely be considered as being in its normal functional environment. The pursuit of sub-picosecond short wavelength pulses to achieve near-atomic resolution of isolated biological macromolecules therefore seems questionable from a number of perspectives, particularly in the face of developments in fluorescence [91, 92] and electron microscopies [93].

The appeal and viability of Bragg's X-ray microscope remains, and modern implementations are foreseeable as powerful XFELs come online [20] and as relevant laboratory-based advances [21, 25] are increasingly recognised [94]. Transient grating IICV SLM detectors would play a key role in such developments. They should also find use in XPCS studies to enable instantaneous optical readout of a detection plane in the interval between two ultrashort pulses of scattered radiation.

### V. SUMMARY

Transient gratings for incoherent ionizing to coherent visible spatial light modulators (IICV SLMs) can foreseeably be implemented using several technologies, including color and F-centers, liquid crystals, ferroelectrics, and cloud and bubble chambers. A few of their relative merits and considerations have been outlined. Generally, the concept is an extremely powerful and useful one, both in the broader context of Bragg's X-ray microscope, and through its potential to allow ultrafast interrogation of an ionizing radiation detection plane or volume. This work outlines a vision as to how it can open new approaches to holographic storage of multidimensional short wavelength ionizing radiation data, as well as sketching possible contemporary ways to reconstruct the phase properties of an ionizing radiation field in the visible optics domain. We have further indicated its potential utility in prospective molecular structure-dynamics developments, along with some associated considerations. It is hoped that these suggestions will stimulate corresponding and further developments in the field of combined ultrafast laser/X-ray science.

# ACKNOWLEDGEMENTS

WKF wishes to acknowledge ARCCoE Coherent X-ray Science grant CE0561787.